\documentclass{emulateapj}

\newcommand{\gr}{$\gamma$-ray \,}
\newcommand{\grs}{$\gamma$-rays \,}
\newcommand{\gsim}{\,\raisebox{0.2em}{$>$}\!\!\!\!\!
\raisebox{-0.25em}{$\sim$}\,}
\newcommand{\lsim}{\,\raisebox{0.2em}{$<$}\!\!\!\!\!
\raisebox{-0.25em}{$\sim$}\,}

\shorttitle{Surplus diffuse Galactic \gr emission}
\shortauthors{V\"olk and Berezhko}

\begin{document}

\title{ON THE {\it FERMI}-LAT SURPLUS OF THE DIFFUSE GALACTIC GAMMA-RAY
  EMISSION}
 

\author{H.J. V\"olk\altaffilmark{1}
    and E.G. Berezhko\altaffilmark{2}
}

\altaffiltext{1}{Max-Planck-Institut f\"ur Kernphysik, P.O. Box 103980, D 69029
  Heidelberg, Germany}             
\altaffiltext{2}{Yu.G. Shafer Institute of Cosmophysical Research and Aeronomy,
                     31 Lenin Ave., 677980 Yakutsk, Russia}     

\email{Heinrich.Voelk@mpi-hd.mpg.de}

\begin{abstract}
  Recent observations of the diffuse Galactic \gr emission (DGE) by the {\it
    Fermi} Large Area Telescope ({\it Fermi}-LAT) have shown significant
  deviations, above a few GeV until about 100 GeV, from DGE models that use the
  GALPROP code for the propagation of cosmic ray (CR) particles outside their
  sources in the Galaxy and their interaction with the target distributions of
  the interstellar gas and radiation fields. The surplus of radiation observed
  is most pronounced in the inner Galaxy, where the concentration of CR sources
  is strongest. The present study investigates this ``{\it Fermi}-LAT Galactic
  Plane Surplus'' by estimating the \gr emission from the sources themselves,
  which is disregarded in the above DGE models. It is shown that indeed the
  expected hard spectrum of CRs, still confined in their sources (SCRs), can
  explain this surplus. The method is based on earlier studies regarding the
  so-called EGRET GeV excess which by now is generally interpreted as an
  instrumental effect. The contribution from SCRs is predicted to increasingly
  exceed the DGE models also above 100 GeV, up to \gr energies of about ten
  TeV, where the corresponding surplus exceeds the hadronic part of the DGE by
  about one order of magnitude. Above such energies the emission surplus should
  decrease again with energy due to the finite life-time of the assumed
  supernova remnant sources. Observations of the DGE in the inner Galaxy at 15
  TeV with the Milagro \gr detector and, at TeV energies, with the ARGO-YBJ
  detector are interpreted to provide confirmation of a significant SCR
  contribution to the DGE.

\end{abstract}


\section{Introduction} 

The diffuse Galactic \gr emission (DGE) from the full sky has recently been
analyzed and compared with the observations with the {\it Fermi} Large Area
Telescope ({\it Fermi}-LAT) for high energies (HE; $200~\mathrm{MeV} \leq
\epsilon_{\gamma} \leq 100~\mathrm{GeV}$) \citep{ackermann12}. The DGE had been
modeled using the GALPROP code \citep[e.g.][]{moskalenko98, strong00,
  porter08}. For a review, see \citet{strong07a}. These phenomenological
models were constrained to reproduce directly measured cosmic ray (CR) data and
were then used iteratively to calculate the DGE \citep[e.g.][]{strong04c}. To
construct a model for the expected total \gr emission, the \gr emission from
the resolved point sources together with the residual instrumental \gr
background and the extragalactic diffuse \gr background -- both assumed to be
isotropic \citep{abdo10g} -- were added to the DGE model. In the inner Galaxy,
the emission of the resolved sources apparently reaches a fraction of $\sim
50$~percent of the expected overall spectral energy flux density at
$\epsilon_{\gamma}\approx 100$~GeV \citep{ackermann12}.

These overall emission models describe the {\it Fermi}-LAT data well at high
and intermediate latitudes and thereby show that the so-called EGRET GeV excess
\citep[e.g.][]{hunter97} does not exist in the form previously inferred
\citep{stecker08, abdo09a, ackermann12}. However, in the Galactic Plane these
models systematically underpredict the data above a few GeV, and they do so
increasingly above about 10 GeV until 100 GeV \citep[see Fig. 15
of][]{ackermann12}. In the present paper this difference between data and model
will be called the "{\it Fermi}-LAT Galactic Plane Surplus" (FL-GPS). It is
most pronounced in the inner Galaxy. According to \citet{ackermann12}, it can
however also be seen in the outer Galaxy, with even a small excess at
intermediate latitudes\footnote{An uncommented overprediction of the models can
  be seen in the energy interval between 10 and 100 GeV for the polar cap
  region south in Fig. 13 of the same authors. This apparent discrepancy is
  characterized by a gamma-ray flux that is more than an order of magnitude
  smaller than in the inner Galaxy and varies in an irregular way in
  energy. Due to this aspect of the data, the discrepancy will not be discussed
  in the sequel.}.

The GALPROP code is constrained by the charged energetic particles directly
measured in the neighborhood of the Solar System which are by assumption truly
diffuse CRs. Therefore the above discrepancy is not too surprising, because in
this comparison the \gr emission from particles {\it within} the CR sources is
only taken into account for those \gr sources that are resolved by the
instrument. 

The dominant part of the \gr sources resolved by the {\it Fermi}-LAT, with 1451
items listed in the {\it Fermi}-LAT 1FGL catalog and taken into account in
  the Ackermann et al. (2012) analysis, are Pulsars, as far as the Galaxy is
concerned.  Except for the Crab Nebula and Vela X the HE \gr emission
from Pulsar Wind Nebulae may actually be again Pulsar radiation, even
  though most recently three more Pulsar Wind Nebulae have been identified with
  {\it Fermi}-LAT \citep{acero13}. For purposes of their \gr emission these
objects are assumed in the present paper to be sources of energetic electrons
and positrons, but not sources of nuclear CRs. Of the latter presumably only a
handful have been resolved, and are thus included in the overall {\it
  Fermi}-LAT emission model \citep{ackermann12}. In all probability the
majority of nuclear CR sources remains unresolved, and is therefore excluded
from that model. As a consequence the FL-GPS can be expected to be a physical,
not an instrumental effect.

Independently of whether they are resolved or not, the nuclear CR sources are
presumably concentrated in the Galactic disk, if they are the consequence of
star formation processes. They are assumed in the present paper to be the
shell-type supernova remnants (SNRs), regardless whether they are isolated or
embedded in stellar associations, e.g. in Superbubbles. The fact that the
FL-GPS is concentrated in the inner Galaxy is then the result of the well-known
concentration of SN explosions in the inner Galaxy \citep[e.g.][]{case98} and
in the inner parts of other galaxies \citep{drag99}. This concentration is
  also confirmed by the Galactic distribution of Pulsars as compact remnants of
  core-collapse SN explosions \citep{lorimer04}. The total \gr emission does
not have such a strong radial gradient in the Galactic Plane, as observed at
comparatively low energies were the purely diffuse emission should dominate, by
e.g. the COS-B satellite for $\epsilon_{\gamma} > 150$~MeV \citep{strong88} and
the EGRET instrument on the CGRO satellite for $100~\mathrm{MeV} <
\epsilon_{\gamma} < 10^4$~MeV \citep[][]{strong96,digel96}. This difference has
also been discussed by \citet{paul01}.

A weak gradient of the diffuse emission has been interpreted theoretically as
the consequence of preferential (faster) convective CR removal from the disk
into the halo in the inner Galaxy, where the higher CR source density and the
decrease of the Galactic escape velocity with increasing Galactic radius drive
a faster {\it Galactic Wind} \citep{breitschwerdt02}. This is a nonlinear
propagation effect. Therefore the concentration of the FL-GPS in the inner
Galaxy is largely the result of the radial gradient in the CR source density,
because the diffuse CR density is largely independent of radius\footnote{A
  physically different interpretation has been advanced in terms of a
  compensatory radial increase of the $W_\mathrm{CO}$-to-$N({\mathrm{H}}_2)$
  scaling factor, regarding the gas target for the gamma-ray emission, on
  account of a radial decrease of the metallicity in the Galactic disk inferred
  from observations in external galaxies \citep{strong04c}.}.

The dependence of the FL-GPS on \gr energy is another aspect which is suggested
to be due to the difference between the diffuse particle spectra and the
particle source spectra. In a selfconsistent model for energetic particle
propagation in such a Galactic Wind \citep{pvzb97}, where nonlinear damping of
the scattering magnetic irregularities balances their growth due to the outward
CR streaming, this spectral difference is naturally explained.

The theoretical interpretation of the location of the FL-GPS in the Galaxy and
of its energy dependence, presented here, is therefore entirely based on the
{\it propagation characteristics of the diffuse CR population in the Galaxy},
both in its dependence on the radial distance from the axis of rotation as well
as in its variation with particle energy.

From a purely phenomenological point of view it is only necessary to assume
that the DGE is essentially independent of radius in the Galactic Plane and
that the diffuse energy distribution of the charged energetic particles is
significantly softer than the energy distribution in their sources. Indeed,
starting with the work of \citet{juliusson72} on the decrease of the flux ratio
of secondary to primary nuclear particles with increasing rigidity, as observed
in the neighborhood of the Solar System, the particle energy spectra inside
these sources have been inferred to be substantially harder than the energy
spectrum of the average CRs in the Interstellar Medium (ISM) \citep[see
e.g.][for a recent review]{ptuskin12}.  Even though the volume fraction of the
active sources is presumably quite small, the energy density of the energetic
particles there is very high, coexisting together with a thermal gas density
comparable to that in the average ISM. In this context an active source is
meant to be one which still basically contains the particles it has
accelerated. These particles shall be called source cosmic rays (SCRs). As
demonstrated for the first time in \citet{bv00}, the contribution of SCRs to
the DGE from accelerated CRs - still confined in their parent SNRs -
progressively increases with energy due to their hard spectrum. Therefore this
contribution becomes unavoidably significant the latest at VHE energies. In a
subsequent paper the SCR contribution was studied in more detail
\citep{bv04}. For the following these papers will be referred to as BV00 and
BV04, respectively. A later study of the contribution of unresolved source
populations has been made by \citet{strong07b}.

Shell-type SNRs have been observed, especially at very
high energy (VHE; $\epsilon_{\gamma} \geq 100$~GeV), to be important sources of
\grs \citep[e.g.][]{hinton09}. It is less clear, in which phase of their
lifetime they lose most of the accelerated particles. This is determined by a
complex interplay of the dissipation of magnetic field fluctuations inside
\citep[e.g.][]{pz03,pz05} and the hydrodynamic break-up of the overall
expanding SNR shell.

The shell-type SNRs are not the only VHE \gr sources in the Galaxy. In fact,
the most abundant resolved Galactic \gr sources at VHE are Pulsar Wind Nebulae
(PWN) \citep[e.g.][]{deona13}. These are in all probability the result of the
winds from young, energetic Pulsars with characteristic ages $\lsim 10^5$~yrs,
and of the accelerated population of electrons in them which, in particular,
emit inverse Compton (IC) \grs in an almost calorimetric fashion
\citep{deona13}. The particle acceleration and resulting emission is quite
complex and poorly understood. PWN observations, including the radio and X-ray
range, are often fitted by a phenomenological model going back to
\citet{aharonian99} which assumes an energy distribution of the radiating
electrons in the form of a power law with a cutoff, in a uniform magnetic field
and in a radiation field approximately equal to the Cosmic Microwave Background
(CMB) \citep[e.g.][]{abramowski12}. The \gr spectral energy densities of PWNs
are therefore assumed to be typically peaked with a cutoff at a few TeV. The
lifetime $\tau_\mathrm{PWN}$ of TeV emitting leptons in typical PWNs has been
estimated to be $\sim 40$~kyr \citep{dejager09}. In a rough sense the total
number $N_\mathrm{PWN}$ of PWNs in the Galaxy is then $N_\mathrm{PWN} \sim 800
(\tau_\mathrm{PWN}/40 \mathrm{kyr}) (\nu_\mathrm{PWN}/2)$, where
$\nu_\mathrm{PWN}$ denotes the Galactic VHE-emitting PWN rate in units of core
collapse events per century \citep{deona13}. This is at least as large as the
number of active shell-type SNRs (see below). As a result the population of
unresolved PWNs may contribute to the FL-GPS, and substantially to the
  total VHE emission, even though this is difficult to estimate for the
reasons given above.

The relative size of the contributions of the unresolved parts of the various
\gr source populations is an important question. Presumably at energies beyond
about 10 TeV the PWN fraction is small due to the cutoff in the electron
spectra as a result of radiative losses. The spectra of nuclear particles in
shell-type SNRs can, on the other hand, extend to energies of many hundreds of
TeV with differential energy spectra close to a power law $\epsilon^{-\gamma}$,
where $2.0 \lsim \gamma \lsim 2.2$. Such an extent in energy is in particular
expected in young objects, like the type Ia objects SN1006 \citep{bkv12} or
Tycho's SNR \citep{bkv13}, where the escape of the highest-energy particles is
arguably still weak. In very energetic, young core collapse supernovae, \gr
energies well in excess of 100 TeV might be reached \citep{bell13}.  However,
the corresponding Supernova rate is expected to be so low \citep{pzs13} (less
than about 1\% of the total Supernova rate) that their contribution is
disregarded here.

In this paper only the contribution of unresolved shell-type SNRs to the
Galactic \gr emission will be discussed in detail, following the earlier work
by BV00 and BV04. As will be shown, the radiation from these SCRs which extends
with a hard power-law-type energy spectrum to multi-TeV \gr energies can
readily explain the above-mentioned FL-GPS and predicts its monotonic increase
from $\epsilon_{\gamma}\sim 100$~GeV into the multi-TeV region. The evaluation
of the contribution of the unresolved PWN population is left for a future
study.

\section{Extrapolation of the expected overall {\it Fermi}-LAT emission model beyond 100 GeV}

The {\it Fermi}-LAT overall emission model corresponds to a spectral energy
distribution (SED) $I_\mathrm{tot} = I_\mathrm{DGE} + I_\mathrm{RS} +
I_\mathrm{IB}$, where $I_\mathrm{DGE}$ denotes the sum of the truly diffuse
hadronic plus inverse Compton plus Bremsstrahlung emission spectra,
$I_\mathrm{RS}$ is the contribution from the resolved sources, and
$I_\mathrm{IB}$ corresponds to that of the isotropic backgrounds.  The resolved
sources -- which include few shell-type SNRs, many Pulsars, and, at energies
$\epsilon_{\gamma}\gsim 100$~GeV, the resolved PWNs -- are expected to have a
harder spectrum than the DGE.  Therefore $I_\mathrm{tot}$ corresponds to a
harder spectrum than the approximate $\epsilon_{\gamma}^{-2.75}$-dependence of
the modeled DGE in the inner Galaxy at energies $\epsilon_{\gamma}\gsim 10$~GeV
below any cutoff. To a large degree this is due to the contribution of the IC
component and especially, that of the resolved sources.
          
Given the representation by the Fermi-LAT collaboration, the most reasonable
initial extension beyond 100 GeV of $I_\mathrm{tot}$ appears to be an
extrapolation by a power-law distribution $I_\mathrm{tot}\propto
\epsilon_{\gamma}^{-\gamma}$ of the same spectral index $\gamma$ which it has
between $\epsilon_{\gamma}\sim 10$~GeV and $100$~GeV.  However, the \gr
spectral energy densities of PWNs are assumed to be typically peaked with a
cutoff at several TeV. The truly diffuse leptonic emission is expected to cut
off at similar energies. Therefore the component
$I_\mathrm{tot}-I^{\pi}_\mathrm{GCR}$ is multiplied by an exponential term
$\exp(-\epsilon_{\gamma}/1~\mbox{TeV})$.  $I_\mathrm{tot}$ extrapolated in this
form is shown by the dashed curve in Fig.\ref{fig1}\footnote{Admittedly, this
  procedure disregards other possible resolved sources, like \gr binaries,
  which may reach significantly higher \gr energies beyond 10 TeV
  \citep[e.g.][]{aha06}.}. From its construction $I_\mathrm{tot}$ is an
approximate lower limit to the expected SED without unresolved sources.

The program of the present paper is then to calculate the contribution of the
unresolved shell-type SNRs in order to predict a lower limit to the total
emission in the inner Galaxy in the GeV range and above. As will be shown, this
lower limit implies an increasing discrepancy between measured and expected
``diffuse Galactic emission' in the region of few GeV $< \epsilon_{\gamma}<$
100~GeV, and thus can explain the FL-GPS. The discrepancy is predicted to
increase up to energies $\sim 10$~TeV in the VHE range. Beyond that \gr energy
the contribution of the unresolved shell-type SNRs goes down due to energy
dependent escape of CRs from shell-type SNRs that makes for a shorter
confinement time of CRs with higher energy.

\section{Results and discussion}
\begin{figure*}
  \plotone{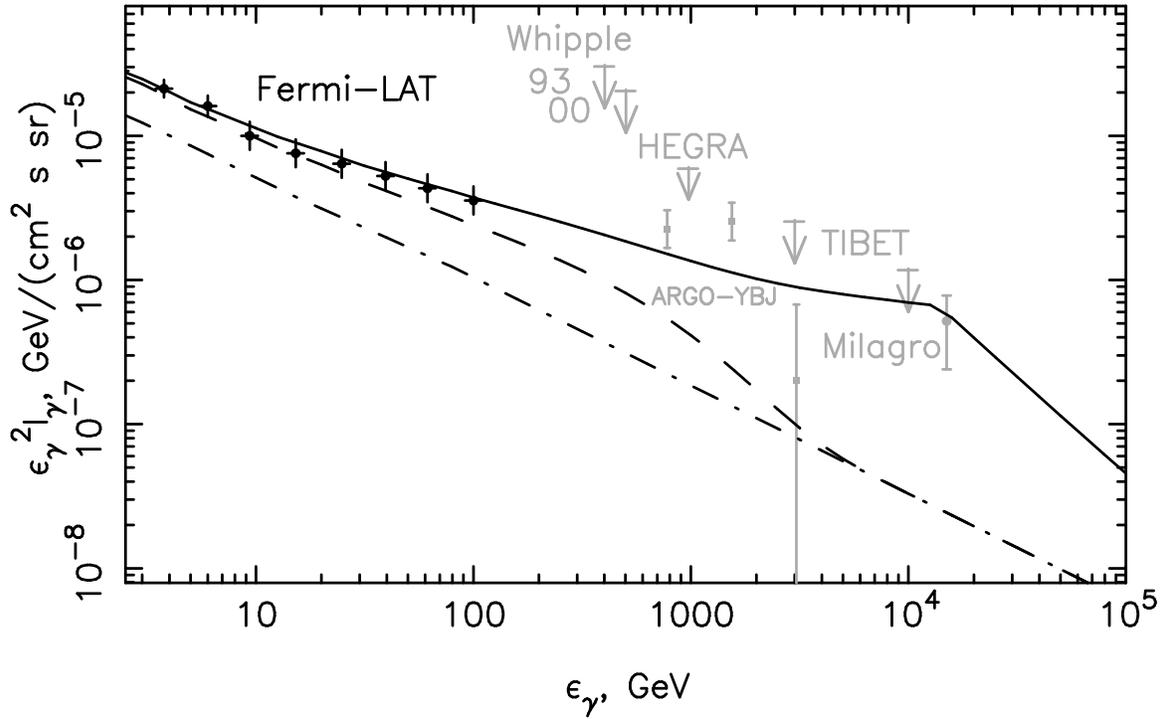} \figcaption{ The diffuse \gr spectrum of the low
    latitude inner Galaxy ($-80^{\circ}<l<80^{\circ}$, $|b|\leq8^{\circ}$).
    The expected $\pi^0$-decay \gr component $I_\mathrm{GCR}^{\pi}$ created by
    truly diffuse GCRs in the Galactic Disk is shown by the dash-dotted
    line. The total expected emission $I_\mathrm{tot}$ (= modeled diffuse
    Galactic emission $I_\mathrm{DGE}$ plus detected sources and isotropic
    backgrounds) up to $\epsilon_{\gamma} = 100$~GeV \citep{ackermann12}
    corresponds to the dashed line. This $I_\mathrm{tot}$ is extrapolated
    beyond 100 GeV by a power law with an assumed cutoff in
    $I_\mathrm{tot}-I_\mathrm{GCR}^{\pi}$ (see text). The solid line represent
    the total expected emission including the contribution of SCRs confined
    inside unresolved SNRs. The {\it Fermi}-LAT data \citep{ackermann12}
      are shown in black color. ARGO-YBJ data in the TeV range \citep{ma11},
    Milagro data at $\epsilon_{\gamma}=15$~TeV \citep{milagro}, the Wipple
    upper limits \citep{reyn93,leboh00}, the HEGRA upper limit \citep{aha01},
    and the Tibet upper limits \citep{tibet} at $\epsilon_{\gamma}=3$ and
    10~TeV, are shown in grey color, see text.  The vertical error bars of
    the {\it Fermi}-LAT data points correspond to the thickness of the gray
    region which includes systematic error in Fig. 15 of \citet{ackermann12}.
\label{fig1} }
\end{figure*}

In Fig.\ref{fig1} a calculated \gr spectrum
$I_{\gamma}=I_\mathrm{tot}+I_\mathrm{SCR}$ of the low latitude inner Galaxy
($-80^{\circ}<l<80^{\circ}$, $|b|\leq8^{\circ}$) is presented. Besides the
overall emission model $I_\mathrm{tot}$, discussed in the previous section, 
it includes the contribution $I_\mathrm{SCR}$ of SCRs
confined within unresolved SNRs. The SCR contribution is calculated based on
the approach developed in  BV04:
\[
I_\mathrm{SCR}=I_\mathrm{GCR}^{\pi}R(1+R_\mathrm{ep}),
\]
where $R$ depends on the Galactocentric distance as a result of the non-uniform
distribution of the Galactic SNRs (see Introduction).  $I_\mathrm{SCR}$
comprises the $\pi^0$-decay component due to hadronic SCRs
$I_\mathrm{SCR}^{\pi}=I_\mathrm{GCR}^{\pi}R$ and the inverse Compton (IC)
component due to electron SCRs
$I_\mathrm{SCR}^\mathrm{IC}=I_\mathrm{GCR}^{\pi}RR_\mathrm{ep}$.  Here
$I_\mathrm{GCR}^{\pi}$ is chosen as the $\pi^0$-decay component due to hadronic
GCRs. At \gr energy $\epsilon_{\gamma}< 100$~GeV the quantity
$I_\mathrm{GCR}^{\pi}(\epsilon_{\gamma})$, determined by \citet{ackermann12},
and its power law extrapolation $I_\mathrm{GCR}^{\pi}\propto
\epsilon_{\gamma}^{-2.75}$ to higher energies $\epsilon_{\gamma}> 100$~GeV is
used.

The dimensionless ratio $R=I_\mathrm{SCR}^{\pi}/I_\mathrm{GCR}^{\pi}$ according
to BV04 has the form
\begin{equation} R(\epsilon_{\gamma})=0.07 
\frac{N_\mathrm{g}^\mathrm{SCR}}{N_\mathrm{g}^\mathrm{GCR}}
\left(\frac{T_\mathrm{p}}{10^5~\mbox{yr}}\right)
\left(\frac{\epsilon_{\gamma}}{1~\mbox{GeV}}\right)^{0.6},
\label{eq2}
\end{equation} 
where $N_\mathrm{g}^\mathrm{GCR}$ and $N_\mathrm{g}^\mathrm{SCR}$ are the gas
number density in the Galactic disk and inside SNRs, respectively. The first
factor in this expression,
$N_\mathrm{g}^\mathrm{SCR}/N_\mathrm{g}^\mathrm{GCR}$, represents the ratio of
target nuclei for the two populations of CRs. The second factor,
$T_\mathrm{p}$, represents the fact that the number of SNRs, wich confine SCRs
with energy $\epsilon$, is proportional to $T_\mathrm{p}(\epsilon)$.  The third
factor $\epsilon_{\gamma}^{0.6}$ is due to the harder SCR spectrum compared
with the GCR spectrum.

Contrary to the earlier considerations of BV00 and BV04, where the interior SNR
magnetic field was assumed to be time independent, the proton confinement time
$T_\mathrm{p}(\epsilon \approx 10\epsilon_{\gamma})$ inside the expanding SNR
is determined taking into account magnetic field amplification and its time
dependence $B\propto (\rho V_\mathrm{S}^2)^{1/2}\propto t^{-3/5}$. Magnetic
field amplification in strong shocks is a general theoretical concept arising
from the expected nonlinear growth of unstable magnetic irregularities produced
by the accelerating CRs themselves \citep{bell04}. Field amplification is
however also clearly observed in a number of SNR sources
\citep[e.g.][]{vbk05}. Here it is assumed to be a universal effect in
shell-type SNRs \citep[e.g.][]{bell13}. In this case the maximum energy of SCR
protons, accelerated at a given SNR evolutionary epoch, decreases with time
according to the relation $\epsilon\propto BR_\mathrm{S}V_\mathrm{S}\propto
t^{-4/5}$.  This leads to the following expression for the proton confinement
time
\begin{equation}
T_\mathrm{p}=\mbox{min}\{T_\mathrm{SN},t_0(\epsilon/\epsilon_\mathrm{max})^{-5/4}~\mbox{yr}\},
\label{eq3}
\end{equation} 
where $T_\mathrm{SN}$ is the time until which SNRs can confine the main
fraction of accelerated particles, $\epsilon_\mathrm{max}$ is the maximal
energy of protons that can be accelerated in SNRs, achieved at the beginning of
the Sedov phase $t=t_0$.  The decrease of the proton confinement time
$T_\mathrm{p}$ with energy $\epsilon$ is due to the diminishing ability of the
SNR shock to produce high-energy CRs during the Sedov phase, which implies the
escape of the highest energy CRs from the SNR \citep{bk88,ber96,byk96}.
Therefore the description in terms of the escape time $T_\mathrm{p}$ is
equivalent to an initial production of the SCR spectrum $N_\mathrm{SCR}\propto
\epsilon^{- \gamma}$ for $mc^2 < \epsilon < \epsilon_\mathrm{max}$ in the
evolving SNR, following the SN explosion proper. Subsequently this spectrum is
confined over the time-scale $T_\mathrm{p}(\epsilon)$. This weighs the fraction
of the highest energies $\epsilon_\mathrm{max}$ with the Sedov time $t_0$,
lower-energy particles with a larger time that increases with decreasing
$\epsilon$, and ``low-energy'' particles with $T_\mathrm{SN}$.

Note that, due to the considerably harder SCR energy spectrum compared with the
GCR spectrum, the factor $R\propto \epsilon_{\gamma}^{0.6}$ grows with energy
up to a maximal energy $\epsilon_{\gamma}^*\approx
0.1\epsilon_\mathrm{max}(t_0/T_\mathrm{SN})^{4/5}$ of \grs produced by SCRs
confined in SNRs for the whole active evolutionary time
$T_\mathrm{p}=T_\mathrm{SN}$. The spectrum $I_\mathrm{SCR}(\epsilon_{\gamma})$
at $\epsilon_{\gamma}<\epsilon_{\gamma}^*$ is predominantly produced by SNRs of
ages $t \approx T_\mathrm{SN}$. Therefore it is only weakly sensitive to the
details of SNR evolution at the epochs $t< T_\mathrm{SN}$.  At higher energies
$\epsilon_{\gamma}>\epsilon_{\gamma}^*$, the ratio $R\propto
\epsilon_{\gamma}^{0.6-5/4}$ goes down with energy and the SCR contribution
decreases due to the decrease of the confinement time $T_\mathrm{p}$.

The ratio $R_\mathrm{ep}=I_\mathrm{SCR}^\mathrm{IC}/I_\mathrm{SCR}^{\pi}$ is
calculated as described in BV04 with a time-independent interior magnetic field
value $B=30$~$\mu$G.

Since the dominant part of SNRs in our Galaxy is due to type II SNe with a
relatively low progenitor star mass, which do not form extended wind bubbles
during their evolution, here only the contribution of SNRs expanding into a
uniform ISM will be considered.

The calculations presented in Fig.\ref{fig1} have been performed with the
following values of the relevant physical parameters: sweep up time
$t_0=10^3$~yr, confinement time of SCRs inside SNRs $T_\mathrm{SN}=3\times
10^4$~yr, electron to proton ratio $K_\mathrm{ep}=10^{-2}$, effective SCR power
law index $\gamma=2.15$, and 10 percent efficiency of SCR production in
SNRs\footnote{Although theoretical calculations lead to a harder particle
  spectrum, $\gamma\leq 2.0$, the choice of an effective $\gamma=2.15$ for the
  SCR {\gr}-production allows the approximate inclusion of HE \gr emission from
  localised strong gas density enhancements overrun by the SNR shock, as one
  might interpret SNR-associated \gr emissions observed, limited largely to the
  HE range \citep[e.g.][]{ackermann13,giordano12,bkv13}.}.  Since the SCRs are
confined inside the SNRs within a thin shell behind the shock, where also the
shock-compressed ISM gas is situated, the interior SNR gas number density
$N_\mathrm{g}^\mathrm{SCR}=4N_\mathrm{g}^\mathrm{GCR}$ is used, where
$N_\mathrm{g}^\mathrm{GCR}$ is the ISM gas number density. Since during the
last ten years convincing observational evidence of considerable SNR magnetic
field amplification \citep{vbk05} leading to the production of CRs with energy
up to $\epsilon_\mathrm{max}\approx 2\times 10^6$~GeV \citep{bv07} has been
found, the value $\epsilon_\mathrm{max}=2\times 10^6$~GeV is used here. It is
considerably larger compared with BV00 and BV04.

Due to the cutoff introduced for the component
$I_\mathrm{tot}-I_\mathrm{GCR}^{\pi}$ the calculated SED
$\epsilon_{\gamma}^2I_{\gamma}$ represents a lower limit for the expected \gr
background at energies $\epsilon_{\gamma} > 1$~TeV. It should nevertheless be
emphasized here that $I_\mathrm{SCR}/I_\mathrm{GCR}^{\pi}$ is given in terms of
the ratios $I_\mathrm{SCR}^{\pi}/I_\mathrm{GCR}^{\pi}$ and
$I_\mathrm{SCR}^\mathrm{IC}/I_\mathrm{GCR}^{\pi}$, and that
$I_\mathrm{GCR}^{\pi}$ is determined by the diffuse nuclear CR distribution
which is rather well-known at least until the ``knee'' of the all-particle GCR
spectrum. Therefore $I_\mathrm{SCR}$ does not depend on the form of the
extrapolation of $I_\mathrm{tot}$ beyond 100 GeV.

As demonstrated in BV04 there are two physical parameters which significantly
influence the expected \gr emission from the ensemble of shell-type SNRs: the
CR confinement time $T_\mathrm{SN}$ and the mean magnetic field strength $B$
inside the SNRs. For a conventional value $B=10$~$\mu$G the expected \gr flux
from SNRs exceeds the HEGRA upper limit considerably if the SCR confinement
time is as large as $10^5$~yr. The contradiction can be resolved either if one
suggests an appreciably higher postshock magnetic field $B\gsim 30$~$\mu$G or
if the SCR confinement time is as small as $T_\mathrm{SN}\sim 10^4$~yr. These
possibilities can be attributed to field amplification by the SCRs
themselves. In fact, nonlinear field amplification may also lead to a
substantial decrease of the SCR confinement time: according to \citet{pz03}
maximal turbulent Alfv$\acute{e}$n wave damping with its corresponding increase
of CR mobility could make the SCR confinement time as small as
$T_\mathrm{SN}\sim 10^4$~yr.

The only physical parameter which strongly influences the IC \gr production
rate for given synchrotron emission rate is the SNR magnetic field $B$:
$I_\mathrm{SCR}^\mathrm{IC}\propto B^{-2}$.  Due to the high magnetic field
$B=30$~$\mu$G the overall contribution of the hadronic SCRs to the \gr flux
dominates over that of the electron SCRs: at TeV-energies
$I_\mathrm{SCR}^\mathrm{IC}\approx 0.3I_\mathrm{SCR}^{\pi}$.

As shown in Fig.\ref{fig1}, the discrepancy between the observed ``diffuse''
intensity and standard model predictions at energies above a few GeV
\citep{ackermann12} can be attributed to the SCR contribution alone up to $\sim
100$~GeV.

Unfortunately, existing VHE measurements do not coincide well in their
longitude (l) and latitude (b) coverage with the corresponding ranges for {\it
  Fermi}-LAT that correspond to $-80^{\circ}<l<+80^{\circ}$ and
$-8^{\circ}<b<+8^{\circ}$. Therefore the following comparisons must be taken
with reservations, and for this reason the observational results from these
measurements are indicated in grey color in Fig.\ref{fig1} 1.: As one can see,
the expected \gr flux at $\epsilon_{\gamma}=1$~TeV is consistent with the HEGRA
upper limit, the Whipple upper limits \citep{reyn93,leboh00}, the Tibet-AS
upper limits \citep{tibet}, and with the fluxes measured by the Milagro
detector \citep{milagro} at $\epsilon_{\gamma}=15$~TeV, as well as with those
obtained with the ARGO-YBJ detector \citep{ma11} in the TeV range, taken at
face value\footnote{Even though the Milagro flux refers to a smaller longitude
  interval than used in the present paper, it also refers to a smaller latitude
  range, so that these effects tend to cancel to lowest order.  See however the
  above general reservations on the comparability of these data.}.
\section{Conclusions} 
These considerations demonstrate that the SCRs inevitably make a strong
contribution to the ``diffuse'' \gr flux from the Galactic disk at all energies
above a few GeV, if the population of shell-type SNRs is the main source of the
GCRs. This explains the {\it Fermi}-LAT Galactic Plane Surplus. The
quantitative estimates show in addition that the SCR contribution dominates
over the extrapolated Fermi-LAT model for the total diffuse emission - which
includes detected sources and isotropic Backgrounds - at energies greater than
100~GeV due to its substantially harder spectrum. The diffuse emission measured
by Milagro at 15~TeV and by ARGO-YBJ at TeV energies provide limited evidence
for that.  It will be interesting to see if, and then to which extent, TeV
observations will show an even higher total emission than the one calculated
here. If this was the case it is suggested that such a difference should be
attributed to the emission from Pulsar Wind Nebulae, or even to other particle
sources, some of which were mentioned above. A full consideration of the
potential particle sources is beyond the scope of this paper.

The Galactocentric variation of any surplus over the purely diffuse emission
above a few GeV should correspond to the Galactocentric variation of the ratio
$R$ between the $\pi^0$-decay components of source cosmic rays and truly
diffuse Galactic cosmic rays as a result of the observed SNR distribution. This
implies a concentration in the inner Galaxy.



\acknowledgements 
The authors would like to thank the referee, Olaf Reimer, for a number of
critical and insightful comments that helped to improve the manuscript. This
work has been supported in part by the Department of Federal Target Programs
and Projects (Grant 8404), by the Russian Foundation for Basic Research (grants
13-02-00943 and 13-02-12036) and by the Council of the President of the Russian
Federation for Support of Young Scientists and Leading Scientific Schools
(project NSh-1741.2012.2).

\end{document}